\documentclass[10pt,preprint]{emulateapj}

\usepackage{stmaryrd}
\usepackage{pifont}
\usepackage{amsmath}
\usepackage{amssymb}
\usepackage{indentfirst}
\usepackage{graphicx}

\graphicspath{figures/}

\slugcomment{}

\shorttitle{Mountain building of solid quark stars} %
\shortauthors{Yang \& Xu}

\begin{document}

\title{Mountain building of solid quark stars}

\author{Haifeng Yang\altaffilmark{1}}
\author{Renxin Xu\altaffilmark{1}}
\altaffiltext{1}{School of Physics and State Key Laboratory of
Nuclear Physics and Technology, Peking University, Beijing 100871,
China.\\
{\em Email:} hfyangpku@gmail.com, r.x.xu@pku.edu.cn}

\begin{abstract}
One of the key differences between normal neutron and (bare) quark stars is relevant to the fact that the former are gravitationally bound while the latter self-confined unless their masses approach the maximum mass.
This difference results in the possibility that quark stars could be very low massive whereas neutron stars cannot.
Mountains could also be build on quark stars if realistic cold quark matter is in a solid state, and an alternative estimation of the mountain building is present.
As spinning compact objects with non-axisymmetric mass distribution will radiate gravitational waves, the equations of states of pulsars could be constraint by the amplitude of gravitational waves being dependent on the heights of mountains.
We then estimate the maximum mountains and thus quadrupole moment on
solid quark stars, to be consistent with that by \cite{Owen2005} if
the breaking strain is $\sim 10^{-1}$, addressing that a solid quark
star with mass $<\sim 10^{-2}M_{\odot}$ could be ``potato-like''.
We estimate the gravitational wave amplitude of solid quark stars
with realistic mountains to be order of $h_0\sim 10^{-27}$.

\end{abstract}

\keywords{dense matter --- gravitational waves --- pulsars: general
--- stars: neutron --- elementary particles}

\maketitle

\section{Introduction}
\label{sec:intro} Neutron stars, with mass of $\sim M_\odot$  and
radius of $\sim 10$ km and thus supra-nuclear density, provide as
unique astro-laboratory for testing fundamental theories. Due to
their high mass to radius ratio, neutron stars with strong gravity
could radiate gravitational waves (GWs) in a few potential ways:
inspiralling binaries \citep{Sathyaprakash2009}, supernova core
collapse \citep{New2003,Ott2009}, rotating deformed stars and GW
radiation induced from oscillations and instabilities
\citep{Andersson1998}. See \cite{Andersson2010} for a review of
these topics.

This work is focus on the gravitational waves due to the rotating
deformed stars. It is well-known that neutron stars with
non-axisymmetric distribution of mass may radiate GWs because of the
time-varying quadrupole moment they have. The frequency of the
gravitational waves will be twice of rotation frequency of neutron
stars. As the the frequencies are of $10^{2\sim 3}$ Hz, the pulsars
are good sources of gravitational waves for interferometers
detectors, such as LIGO \citep{Abbott2009} and Virgo
\citep{Acernese2008}. Although LIGO hasn't detect the signal from
pulsars yet, a limitation of the amplitude of the gravitational
waves has been put forward \citep{Abbott2005,Abbott2007,Abbott2010}.

One of the key motivations to detect GWs from pulsar is to provide
constraints on the equation of state of cold matter at supra-nuclear
density, a challenge in understanding the fundamental strong
interaction between quarks.
Pulsars could be neutron or quark stars, even in a solid state
\citep{Xu2003,Xu2009}. Can the LIGO detection be understood in the
regime of solid quark stars? What about GWs if pulsars are really
solid quark star? We are concerned about these issues in this paper.
\cite{Ushomirsky2000} first consider the elastic deforming neutron
stars. Their estimation of the maximum quadrupole moment of the
normal neutron stars is $2.4\times 10^{38}$ g$\cdot$cm$^2$, which
has been updated by \cite{Owen2005} via redefining the shear
modulus. They also give a general equation to calculate the
quadrupole moment induced by shear modulus which can be used to
other stars such as solid quark stars, and \cite{Owen2005} estimates
the quadrupole moment of the solid quark stars to be $2.8\times
10^{41}$ g$\cdot$cm$^2$. The quadrupole moment is also calculated in
other stellar models, such as Hybrid crystalline
clour-superconducting star \citep{Knippel2009,Haskell2007,Lin2007}
and Hybrid and meson condensate stars \citep{Owen2005}.
\cite{Pitkin2011} provides a good review on these.

However, all these studies assume that the equilibrium structure of
the neutron stars is almost spherically symmetric. Certainly, some
solidified mountains would also be possible on the surfaces of sold
quark stars.
Unlike shear-modulus-induced mountains (elastic mountains), we
suggest here latent-heat-induced mountains (solidified mountains)
there. Under this condition, instead of the shear modulus and the
breaking strain, latent heat perform an important role that
determine the height of the mountains. From this point of view, our
model is only one parameter (i.e., latent heat) dependent, rather
than than two parameter (i.e., shear modulus and breaking strain)
dependent in case of elastic mountains.

After estimating the maximum height of the mountains, we find that
if the mass of solid quark stars is small enough that the height is
of the same order with the radius of the star, the star may be
``potato-like''. The critical mass for potato-like solid quark stars
is estimated  to be $\sim 10^{-3\sim -2}M_\odot$, which agrees with
the mutually independent study of \cite{Xu2010}.
Also it is found that, in reality, the actual GW amplitude of a
pulsar would be too small to be detected with LIGO now since the
maximum GW amplitude requires a star to deform into a particular
distribution of mountains.

\section{General arguments}
\label{sec:General}

{\it Mountains on the Earth as an analogy}. Before discussing the
latent-heat-induced mountains on the solid quark stars, we present
an estimation of the maximum height of mountains on the Earth using
the same method.
Consider a mountain with the height $H$ and sectional area $A$.
The earth is assumed to have mass of $M$ and radius $R$.
Local acceleration of gravity is $g$.
If the mountain gets $\delta$ height down, the material at the foot
of the mountain will get energy of $A\delta \rho g$. While this
energy is enough for this amount of material to melt, the mountain
can no longer get higher because the mountain may tend to melt
rather than to keep the height if it is higher than this. Certainly,
this process of melting is somewhat a virtual one.
We can also understand this in another way that this method is
comparing energy between two: a mountain with height $H$ and the
other with height $H-\delta$.
The maximum height of the mountain, $H_{\rm m}$, will then guarantee
an equation of $A\delta\rho gH_{\rm m}=A\delta\rho\Lambda N_A/\mu$,
where $\Lambda N_A$ is the latent heat per mol (with $N_A$ the
Avogadro constant) and $\mu$ is the mass per mol. Therefore we have
$H_{\rm m}=\Lambda N_A/(\mu g)$. On the earth the material that
construct mountains is SiO$_2$, which has $\mu=60$ g/mol and
$\Lambda N_A=854$ kJ/mol. The estimate of the maximum height of
mountains on the earth is thus $14.5$ km. The highest mountain known
on the Earth is Qomolangma with the height of $8.8$ km, which is the
same order of the estimated maximum height.

{\it Mountains on solid quark stars}. In this case, we can also have
the equation of
$$\frac{GM}{R^2}H_{\rm m}A\delta n\chi m_0=A\delta n\Lambda,$$
where $n$ is the number density of the quark clusters, $\chi$ is the
number of quarks in a quark cluster, $m_0=300$ MeV is an assumed
mass of every dressed quark, and $\Lambda$ is the melting energy per
quark cluster. \cite{Michel1988} proposed that quark-alpha particle
with $18$ quarks is possible to exist, while the proton and neutron
are formed with $3$ quarks. We then suggest that the $\chi$ varies
from $3$ to $18$ and choose $\chi=10$ as fiducial value.
\cite{Ishii2007} shows that the nucleon-nucleon potential could be as
depth as $\sim 100$ MeV. The latent heat should thus be considered
to vary from $1$ MeV to $10$ MeV, and $\Lambda = 5$ MeV is chosen as
the fiducial value for furthur calculation. The definition of $A$,
$H_{\rm m}$, $M$ and $R$ is the same as above. One can then has
\begin{equation}\label{hm}
\begin{split}
    H_{\rm m}\simeq & 8 \times 10^3{\rm cm}\left(\frac{R}{10{\rm km}}\right)^2\left(\frac{1.4M_\odot }{M}\right)\\
& \times\left(\frac{\Lambda}{5{\rm
MeV}}\right)\left(\frac{10}{\chi}\right).
\end{split}
\end{equation}
It seems that there are two parameters, $\Lambda$ and $\chi$, but it
is evident that the maximum height is only one parameter dependent,
$\Lambda/\chi$, the average latent heat per quark.

Certainly the stars are not spherical because of many mountains
there. Stars could only be approximately spherical if $H_{\rm m}\ll
R$, otherwise, if $H_{\rm m}/R\sim O(1)$, those gravity-free stars
might be ``potato-like''.
For a low mass solid quark star with then approximately constant
density, $\rho\sim 4\times 10^{14}$g/cm$^3$, one has from
Eq.\eqref{hm}
\begin{equation*}
\begin{split}
\frac{H_{\rm m}}{R}\simeq&0.008\left(\frac{R}{10{\rm km}}\right)\left(\frac{1.4M_\odot }{M}\right)\left(\frac{\Lambda}{5{\rm MeV}}\right)\left(\frac{10}{\chi}\right)\\
=&0.013\left(\frac{10{\rm km}}{R}\right)^2\left(\frac{4\times10^{14}{\rm g/cm}^3}{\rho}\right)\\
&\times\left(\frac{\Lambda}{5{\rm
MeV}}\right)\left(\frac{10}{\chi}\right).
\end{split}
\end{equation*}
When $R\sim \sqrt{0.013}\times10$ km = 1.15 km, we will have $H_{\rm
m}/R\sim O(1)$, and the corresponding mass for potato-like solid
quark stars would be be $M<\sim (\sqrt{0.013})^3\times1.4
M_{\odot}=0.002M_{\odot}$. This result agrees with that of
\cite{Xu2010}, who addressed that quark stars with masses as low as
$10^{-2}\sim10^{-3} M_{\odot}$ are gravity-free.
An important astrophysical consequence of potato-like solid stars
would be of precession with high amplitude, and we expect to test
this by future observations.
Possible low-mass compact stars is a direct consequence of the
suggestion that pulsars could be quark stars, which could be the
central remnants left by the detonation of the accretion-induced
collapse of white dwarfs \citep{Yu2011}.

\section{Gravitational wave radiation from solid quark stars}
\label{sec:gw}

As there are mountains on the stars, these mountains may attribute
to the quadrupole moment of the stars and thus influence the
amplitude of the gravitational waves the stars radiate. However, if
a star accrete mass axisymmetrically, it will have an axisymmetrical
distribution of mountains. Thus, the star will not radiate
gravitational waves at all. It is difficult for us to know the real
distribution of mountains and amplitude of gravitational waves the
star radiate, but we can provide an upper bound of the amplitude, by
assuming a special distribution of mountains on the star. The first
term that attribute to the quadrupole moment is $Y_{22}$, so we can
assume the stars having a height distribution of the mountains as
\begin{equation}\label{HDis}
\begin{split}
h(\theta,\phi)&=\frac{1}{2\sqrt{N}}H_{\rm m}{\rm Re}[Y_{22}(\theta,\phi)]\\
&=\frac{1}{2}H_{\rm m}\sin^2(\theta)\cos(2\phi),
\end{split}
\end{equation}
where $N=15/32\pi$. We give the factor $1/(2\sqrt{N})$ here to
ensure the difference between the maximum value and minimum value is
$H_{\rm m}$. We can define $Q_{22}$ in the way of
\cite{Ushomirsky2000}, $Q_{22}=\int\delta \rho_{22}(r)r^4dr.$ In our
simulation, we simply define a density perturbation of $\delta \rho$
to be of delta function (only to be non zero when $r=R$). By
comparing this with our distribution of the mountains, we can obtain
$Q_{22}=\rho H_{\rm m}R^4/2$. We have thus
\begin{equation}\label{Q22esti}
    \begin{split}
Q_{\rm 22,max}=&4.17\times 10^{42}{\rm g\cdot cm}^2\left(\frac{\rho}{4\times10^{14}{\rm g/cm}^3}\right)\\
&\times \left(\frac{R}{10 {\rm km}}\right)^6
\left(\frac{1.4M_\odot}{M}\right) \left(\frac{\Lambda}{5 {\rm
MeV}}\right)\left(\frac{10}{\chi}\right).
\end{split}
\end{equation}
Let's compare this result with that of \cite{Owen2005}, who proposed
a solid quark star to have a quadrupole moment of $2.8\times
10^{41}$g$\cdot $cm$^2$ by suggesting a breaking strain of
$10^{-2}$.
Our estimation is one order higher than that of \cite{Owen2005}, and
shows that the solidified mountains can provide a bigger quadrupole
moment than elastic mountains do. \cite{Horowitz2009} have done some
work to show that the breaking strain on normal neutron stars can be
as big as $10^{-1}$. Our result about the maximum quadrupole moment
on the solid quark stars would agree with Owen's if the breaking
strain on solid quark stars can also be as high as $10^{-1}$. It is
evident that both the dependence on radius and mass are the same in
this work and that of \cite{Owen2005}.

What can we constrain the equation of state by the GW observations?
To compare the theoretical quadrupole moment with the LIGO S5 data
\citep{Abbott2010}, we are to provide the relationship between
$Q_{22}$ and GW amplitude $h_0$. Assuming a density perturbation of
$\delta \rho={\rm Re}[\rho_{22}Y_{22}(\theta,\phi)]$
\citep{Ushomirsky2000} and following the definition of $Q_{22}$
above, one comes to
\begin{equation}\label{Q22h0}
    h_0=3.87\times 10^{-27}\left(\frac{f}{100{\rm Hz}}\right)^2\left(\frac{1{\rm kpc}}{d}\right)\left(\frac{Q_{22}}{10^{38}{\rm g\cdot
    cm}^2}\right),
\end{equation}
where a factor $1/\sqrt{2}$ before the amplitude of the
trace-reversed perturbation is added for $h_0$ estimation. Similar
treatments can also be found in other gravitational wave literatures
\citep{Jaranowski1998}. The only difference is applying ellipticity
$\epsilon$ rather than quadrupole moment $Q_{22}$. Note the
relationship $\epsilon=\sqrt{8\pi/15}Q_{22}/I_{zz}$, it is easy to
prove that they are equivalent.

With our estimation of the maximum quadrupole moment into the
equation above, we can obtain the maximum amplitude of the
gravitational waves,
\begin{equation}\label{h0esti}
\begin{split}
    h_0^{\rm max}=&1.6\times10^{-22}\left(\frac{f}{100{\rm Hz}}\right)^2\left(\frac{1{\rm kpc}}{d}\right)\left(\frac{\rho}{4\times10^{14}{\rm g/cm}^3}\right)\\
& \times \left(\frac{R}{10\rm
km}\right)^6\left(\frac{1.4M_\odot}{M}\right)
\left(\frac{\Lambda}{5\rm MeV}\right)\left(\frac{10}{\chi}\right).
\end{split}
\end{equation}
LIGO is focusing on the direct detection of gravitational waves.
Recent LIGO S5 data has given an upper bound of the $h_0$ for 116
known pulsars \citep{Abbott2010}. The value of the upper bound of
the 116 pulsars varies from $10^{-26}$ to $10^{-25}$. However, from
Eq.(\ref{h0esti}), it seems that solid quark stars can have
mountains high enough to radiate gravitational waves with amplitudes
as high as $10^{-22}$.
Does this mean that these 116 pulsars cannot be solid quark stars?

Should we believe that the mountains on solid quark stars have the
maximum heights? In addition, in our estimation of
Eq.(\ref{h0esti}), we assume a distribution of $Y_{22}$, which can
attribute most to the gravitational waves.
It is surely quite strange if natural mountains on stars have the
exact distribution of $Y_{22}$ in order to produce the maximum
gravitational waves.

The orogeny in the Earth's crust is powered by the mantle convection
and thus the engagement of tectonic plates. What could be the force
to build mountains on solid quark stars? Elastic energy develops
when a star evolves, and both bulk-invariable and bulk-variable
forces can result in decreases of moment of inertia during a star
quake \citep{Peng2008}. This force would be responsible for mountain
building too.
We can then estimate the real heights of mountains from the glitch
phenomena of pulsars below.
We think glitches occur if the moment of inertia change suddenly
inside a solid quark stars \citep{Zhou2004}. The pulsars' angular
momentum, under the assumption of sphere, is $\sim 3MR^2\Omega/5$.
Making a differential of it, we have $\left|\delta R/R\right|\sim
5\times 10^{-7}[\delta(\Omega/\Omega)/10^{-6}]$. Although this
consideration could not be the real heights of mountains on pulsars,
we think that actually the height of mountains might be of the same
order of $\delta R$. Therefore, replacing $\delta R$ with $H_{\rm
m}$ and inserting this into the estimation of Eq.(\ref{Q22h0}), we
have the estimation of amplitude $h_0$ below,
\begin{equation}\label{h0omega}
\begin{split}
h_0=&1\times 10^{-26}\left(\frac{f}{100{\rm
Hz}}\right)^2\left(\frac{1{\rm kpc}}{d}\right)
\left(\frac{\rho}{4\times10^{14}{\rm g/cm}^3}\right)\\
& \times \left(\frac{R}{10{\rm
km}}\right)^5\left(\frac{\delta\Omega}{\Omega}/10^{-6}\right).
\end{split}
\end{equation}

We draw Fig. 1 to show how $h_0$ varies as the mass of solid quark
stars changes in the model of \cite{Lai2009}, for different glitch
amplitude $\delta\Omega/\Omega$. In the calculations, we choose the
numbers of quarks in a quark clusters to be $18$ and the potential
$U_0$ to be $50$ MeV. We can see that if our estimation of the real
height of mountains is valid, it would be natural that LIGO still
hasn't detect the gravitational waves directly, since the maximum GW
amplitude presented in Eq.(\ref{h0esti}) requires maximum height and
a particular $Y_{22}$-distribution of mountains.
What if the mountains haven't the $Y_{22}$-distribution? This will
be discussed in the next section.

\section{Gravitational waves induced from random distribution}
\label{sec:random}

Above we give an approach to the actual amplitude of GWs using the
glitch phenomenon. However, the amplitude of GWs from quark stars
depends also on another factor. Besides the maximum height of
mountains, the distribution of mountains plays a key role as well.
In the previous section, we give an estimation of the maximum
amplitude of GWs by assuming a specific distribution described in
Eq. \eqref{HDis}. Such a hypothesis is too strong to approach the
physical circumstances. In this section we will discard this
assumption and consider another likely distribution.

It is certainly very difficult to know the real distribution of
mountains on a solid quark star. Nevertheless, an idea comes out
that we can give a random distribution to approach the actual
situation. Yet another problem of the definition of random
distribution emerges. The most strict and physical definition should
satisfy the following requirements. (1) The height of mountains
should varies from 0 to the maximum height of mountains. (2) The
surface of the star should be continuous, say the function
$H(\theta,\phi)$ which describe the distribution of mountains should
be infinitely differentiable. (3) The third requirement comes not
from mathematics, but from physics: the partial derivative of
$H(\theta,\phi)$ should not be too high, otherwise a mountain which
is quite precipitous would tend to fall down.

With all the considerations above, we find it is impossible or at
least very hard to use the most strict definition of random
distribution to approach the real amplitude of GWs. Nonetheless, we
would have a order-of-magnitude estimation of the distribution. The
only thing we do care is the quadrupole moment the distribution
induces. We then use another definition which is not that strict but
we expect the latter definition can give a similar prediction or at
least in the same order.
We simply divide our star into some small pieces, say $100 \times
100$ for example. Each piece has height varying from zero to the
maximum height of mountains in a uniformly random way. This is the
definition of $H_r(\theta,\phi)$ which is discussed below.

As a scatter function defined on sphere, the function of
$H_r(\theta,\phi)$ can be expanded using the spherical harmonics as
following,
$$H_r(\theta,\phi)=\displaystyle{\sum_{l,m}}C_{lm}Y_{lm}(\theta,\phi).$$
Each mode contributes to corresponding mode of GWs. The mode with
the sphere distribution of $Y_{22}(\theta,\phi)$ is believed to
contribute GWs with the biggest amplitude. To pick up this mode we
use the orthogonal property of the sphere harmonics. We multiply the
$H_r(\theta,\phi)$ with ${\rm Re}[{Y_{22}(\theta,\phi)}]$ and
calculate the integral on the sphere. The result is regarded as the
factor before $Y_{22}$ in Eq. \eqref{HDis}, and we can calculate the
quadrupole moment accordingly.

We divide the star into $100\times 100$ pieces, calculate 10000
times in the simulation, take the average of the results and finally
obtain the estimation of the quadrupole moment of the order
$10^{38}{\rm g\cdot cm}^2$. Correspondingly, the amplitude of GWs is
of the order $10^{-27}$. Considering the upper bond given by LIGO S5
data varies from $10^{-26}$ to $10^{-25}$, our results would be good enough to
explain the conflict between maximum estimation and the
observations.

\section{Conclusions and Discussions}
\label{sec:conclusion}

We present a new method of building mountains on solid quark stars.
It is induced by the latent heat rather than shear modulus. The
quadrupole moment brought by this kind of solidified mountains is of
the same order with that of \cite{Owen2005} if one uses $10^{-1}$
(instead of $10^{-2}$) as the fiducial value of the breaking strain
on solid quark stars.

Although solid quark stars potentially radiate strong gravitational
waves, it is necessary to build high mountains with enough energy to
deform the quark stars. It is worth noting that the maximum
amplitude of gravitational waves requires the stars to deform into a
particular distribution of mountains, $Y_{22}$.
With the estimation of reality, we find that the actual amplitude
(with $h_0$ of the order $10^{-27}$) would be too small to be
detected with LIGO now. We expect also gravitational waves being
detected directly in the second or third generation detectors
\citep{Pitkin2011}.


As well as the discussion of gravitational wave radiation, we
predict ``potato-like'' quark stars with mass as low as $<
10^{-2}M_\odot$ (the corresponding stellar radius is smaller than a
few km), since gravity could be negligible there, as in the case of
asteroids.
Another independent reason for potato-like quark stars is shown in
Fig. 1 of \cite{Xu2010}.
Rigid body precesses naturally, but fluid one can hardly. The
observation of a few precession pulsars may suggest a totally solid
state of matter. As discussed above, potato-like quark stars are
gravitationally force-free, and thus free or torque-induced
precession may easily be excited with larger amplitude in low-mass
solid quark stars, which are potato-like.

\section*{Acknowledgments}
This work is supported by the National Natural Science Foundation of
China (Grant Nos. 10935001, 10973002), the National Basic Research
Program of China (Grant No. 2009CB824800), the National Fund for
Fostering Talents of Basic Science (J0630311) and the John Templeton
Foundation.

\clearpage

\begin{figure}
\label{fig:h0-m}
\includegraphics[width=\textwidth]{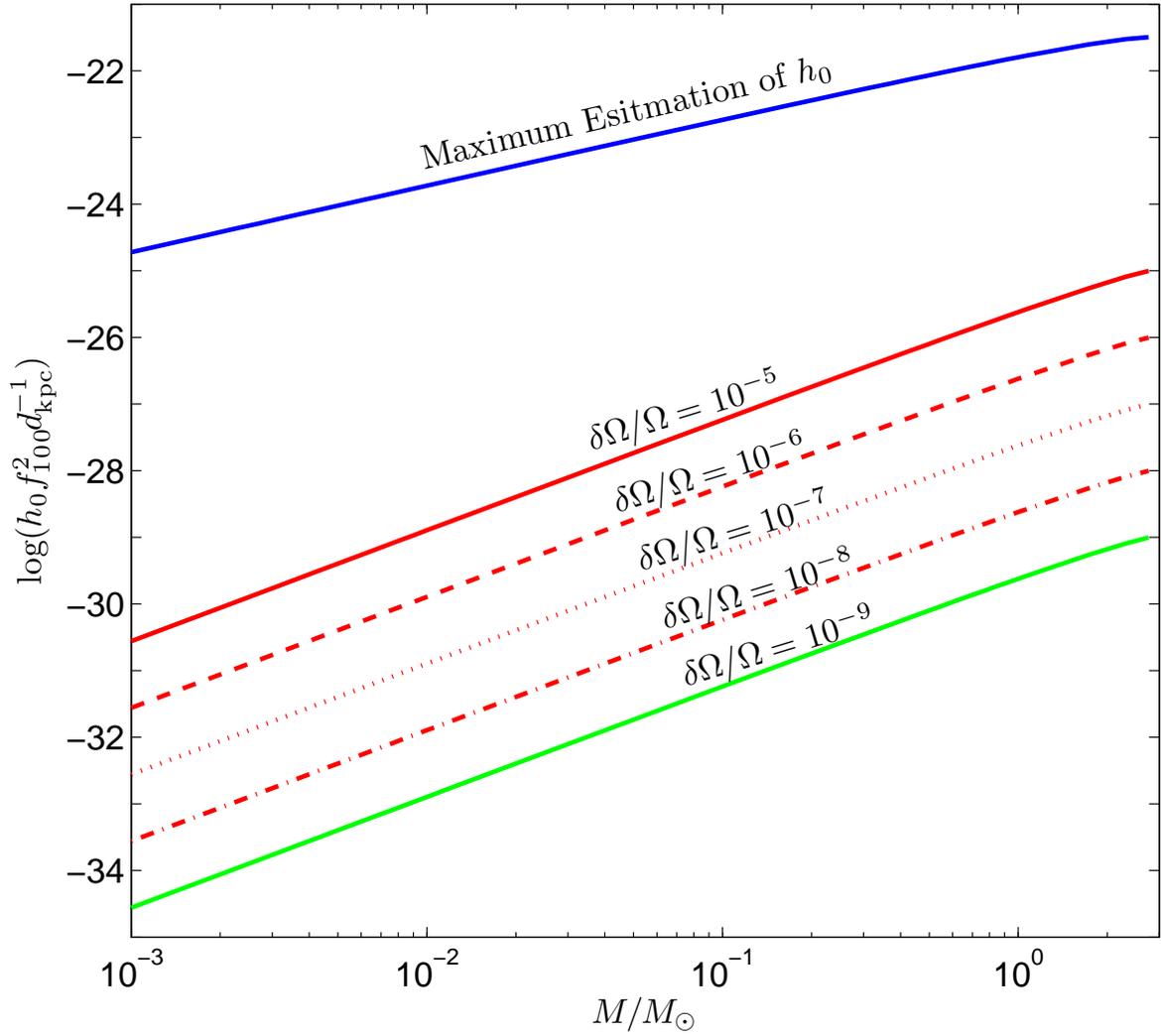}
\caption{Estimation of amplitude of gravitational waves from solid
quark stars in \cite{Lai2009} model ($N_q=18$ and $U_0=50$ MeV).
This figure shows how GW amplitude, $h_0$, varies while the mass of
stars changes. The line on the top is $h_0$ estimated from the
maximum height of mountains with the distribution of $Y_{22}$. The
five lines under it represents for the $h_0$ estimated from the
consideration of glitch. See the text for details.}
\end{figure}

\end{document}